\documentclass{emulateapj}
\usepackage{dsfont}

\newcommand{\myemail}{havenhaus@astro.phys.ethz.ch}

\slugcomment{Accepted for publication in ApJ: May 22, 2014}

\shorttitle{HD100546 multi-epoch scattered light observations}
\shortauthors{Avenhaus et al.}

\begin{document}


\title{HD100546 multi-epoch scattered light observations}

\author{Henning Avenhaus$^{1,2}$, Sascha P. Quanz$^2$, Michael R. Meyer$^2$, Sean D. Brittain$^3$, John S. Carr$^4$, and Joan R. Najita$^5$}
\email{\myemail}

\altaffiltext{1}{Based on observations collected at the European Organisation for Astronomical Research in the Southern Hemisphere, Chile, under program number 090.C-0571(B).}
\altaffiltext{2}{ETH Zurich, Institute for Astronomy, Wolfgang-Pauli-Strasse 27, 8093 Zurich, Switzerland}
\altaffiltext{3}{Department of Physics \& Astronomy, 118 Kinard Laboratory, Clemson University, Clemson, SC 29634, USA}
\altaffiltext{4}{Naval Research Laboratory, Code 7211, Washington, DC 20375, USA}
\altaffiltext{5}{National Optical Astronomy Observatory, 950 N. Cherry Ave., Tucson, AZ 85719, USA}


\begin{abstract}
We present $H$, $K_{\rm s}$ and $L'$ filter polarimetric differential imaging (PDI) data for the transitional disk around HD100546 obtained in 2013, together with an improved re-reduction of previously published 2006 data. We reveal the disk in polarized scattered light in all three filters, achieving an inner working angle of $\sim$0.1\arcsec. Additional, short-exposure observations in the $H$ and $K_{\rm s}$ filter probe the surrounding of the star down to $\sim$0.03\arcsec~($\sim$3 AU). HD100546 is fascinating because of its variety of sub-structures possibly related to forming planets in the disk, and PDI is currently the best technique to image them in the near-IR. Our key results are: (1) For the first time ever, we detect a disk in $L$-band PDI data. (2) We constrain the outer radius of the inner hole to 14$\pm$2 AU and its eccentricity to $<0.133$. (3) We detect a dark lane between $\sim$0.2-0.6$\arcsec$~AU in the front side of the disk, which is likely an effect of the scattering angle and the scattering function of the grains. (4) We find a spiral arm in the northeast which has no obvious connection to spiral arms seen before by other authors further out in the disk, but winds in the same direction (clockwise). (5) The two bright scattering peaks along the semi-major axis are asymmetric, with the southeastern one being significantly brighter. This could be related to the inner companion candidate that is close to the brighter side of the disk at the time of the observations. (6) The scattering color is close to grey between $H$ and $K_{\rm s}$ filter ([$H$]-[$K_{\rm s}$] = 0.19$\pm$0.11), but the scattering in $L'$ filter is significantly weaker ([$H$]-[$L'$] = -1.08$\pm$0.35, [$K_{\rm s}$]-[$L'$] = -1.27$\pm$0.35). (7) We measure the position angle of the disk to be 138$^\circ$$\pm$3$^\circ$, consistent with previous observations. (8) We derive the dust scattering function in the $H$ and $K_{\rm s}$ filter between $\sim$35$^\circ$ and $\sim$130$^\circ$ at two different radii (30-50 and 80-110 AU) and show that our results are consistent with a disk that is more strongly flared in the outer regions.

\end{abstract}



\objectname{HD100546} 


\section{Introduction}
Dozens of circumstellar disks have been successfully resolved in scattered light using high-contrast imaging techniques on large, ground-based telescopes or the Hubble Space Telescope (HST)\footnote{Examples are collected at http://www.circumstellardisks.org/}. A particular powerful technique is polarimetric differential imaging (PDI) which allows for a very accurate subtraction of the central star's point spread function (PSF) revealing the significantly fainter signal of the surrounding disk even without the use of a coronagraph. This gives access to inner working angles as small as $\approx$0.1\arcsec~with 8-m class, ground-based telescopes, which is of great relevance for planet formation studies: At the distance of the observed stars these separations correspond to the innermost few tens of AU of circumstellar disks where most of the planet formation is expected to occur.

Recently, using PDI, numerous circumstellar disks around young, nearby stars were directly imaged at near-infrared (NIR) wavelengths. Interestingly, a lot of these images showed a variety of sub-structures and distinct morphological features in these disks that could be related to planet formation processes, such as gaps, cavities and spiral arms \citep[e.g.][]{avenhaus2014, garufi2013, quanz2013b, grady2013, hashimoto2012, mayama2012, muto2012}. A very interesting target is the young Herbig Ae/Be star HD100546, where first PDI results in the $H$ and  $K_{\rm s}$ filter were presented in \citet{quanz2011}. Basic parameters for this star are given in Table \ref{table:HD142527}. The star is surrounded by a transition disk consisting of a small inner disk between $\sim$0.2--0.7 AU \citep{panic2012}\footnote{Note that studies based on NIR interferometry prefer an outer radius of the inner disk that is larger \citep[$\sim$4 AU,][]{benisty2010, tatulli2011}} followed by a disk gap out to $\sim$13--15 AU and then a large outer disk extending out to a few hundreds of AU \citep[e.g.,][]{pantin2000, augereau2001, grady2001, grady2005, ardila2007}. The disk gap was initially proposed based on SED models \citep{bouwman2003} and observationally confirmed with far UV spectra using HST/STIS \citep{grady2005}. Also the images of the first PDI study of the disk found evidence for a disk rim of the outer disk around $\sim$15 AU \citep{quanz2011}. From ro-vibrational CO emission lines \citet{brittain2009} found evidence for an inner cavity existing not only in the dust but also in the gaseous CO component of the disk \citep[c.f.][]{vanderplas2009}. Prominent, large scale spiral arms were clearly detected in HST images at optical and NIR wavelengths \citep[e.g.,][]{grady2001, ardila2007}. Using ground-based NIR images, \citet{boccaletti2013} found evidence for multiple spiral arms in the southern side of the disk.

\begin{deluxetable}{lcc}
\centering
\tablewidth{0pt}
\tablecaption{Basic parameters of HD100546.
\label{table:HD142527}}           
\tablehead{
\colhead{Parameter} & \colhead{Value for HD100546} & \colhead{Reference\tablenotemark{a}}
}
\startdata
RA (J2000) & 11$^h$33$^m$25$^s$.44  & (1) \\
DEC (J2000) & -70$^\circ$11$'$41$''$.24   & (1)\\
$J$ [mag] & 6.43$\pm 0.02$  & (2)\\
$H$ [mag]& 5.96$\pm 0.03$   &(2)\\
$K_{\rm s}$ [mag]& 5.42$\pm 0.02$  & (2)\\
$L$ [mag]& 4.02$\pm 0.06$  & (3)\\
Mass [M$_\sun$] & 2.4$\pm$0.1 & (4)\\
Age [Myr]& $5...>10 Mry$ & (4),(5) \\
Distance [pc] & $97^{+4}_{-4}$ & (6)\\
Sp. Type & B9Vne & (7) \\

\enddata
\tablenotetext{a}{References --- (1) \citet{perryman1997} (2) 2MASS point source catalog \citep{cutri2003}, (3) \citep{dewinter2001}, (4) \citet{vandenancker1997}, (5) \citet{guimaraes2006}, (6) \citet{vanleeuwen2007}, (7) \citet{houk1975}.
}
\end{deluxetable}

In particular the disk gap was often seen as possible indication of young planets orbiting in the disk \citep[e.g.,][]{bouwman2003, tatulli2011}. Observational support for a companion in the gap was provided by \citet{acke2006} based on temporal changes in the [OI] line profile possibly being a signpost for a yet unseen 20 Jupiter mass planet orbiting within the gap. More recently, \citet{liskowsky2012} observed asymmetric line profiles in the OH spectrum of HD100546 which are consistent with emission coming from an eccentric annulus near the disk rim possibly driven by an orbiting companion. A more direct indication of a close-in companion comes from non-axisymmetric structures in the gaseous CO emission \citep{brittain2013}. The spectro-astrometric signal in the $\nu =1-0$ CO emission varies significantly over a baseline of several years, and can be fit with emission from a non-varying circumstellar disk plus a compact source of emission that varies in velocity as it orbits the star \citep{brittain2013}. The required emitting area ($\sim$0.1 AU$^2$) of the orbiting component can be explained by a circumplanetary disk in agreement with model predictions \citep[e.g.,][]{ayliffe2009}. A first direct upper limit on possible companions inside the gap was provided by \citet{grady2005} who could exclude a stellar companion. Recently, \citet{mulders2013b} used hydrodynamical simulations to model the rounded-off shape of the outer disk rim, which is constrained by mid-infrared (MIR) interferometric data \citep{panic2012}. The apparent gradient in the rim's surface density depends on the disk viscosity and also on the mass of the body orbiting in the gap. These simulations suggested that the mass of the orbiting body is in the range of 60$^{+20}_{-40}$ Jupiter masses.

In addition to the suspected object orbiting in the disk gap, a second planet candidate was discovered by means of high-contrast, direct imaging \citep{quanz2013a}. Using the APP coronagraph installed at VLT/NACO \citep{kenworthy2010}, an $L'$ emission source located roughly $\sim$0.5\arcsec~(de-projected 70 AU) from the central star was detected, i.e., right in the middle of the optically thick, large outer disk. This emission source was best explained with a combination of a point source component and some extended emission, and given its brightness and small separation from HD100546 it is unlikely to be a background object. \citet{quanz2013a} argued that the object is possibly a young, forming gas giant planet that still undergoes gas accretion. This could explain both the observed luminosity (part of the luminosity is coming from the accretion process via a circumplanetary disk) and the apparently smooth circumstellar disk at these separations (the object is young, not yet very massive and hence did not alter the circumstellar disk structure significantly).

The previous paragraphs strongly emphasize that HD100546 is not only an extremely well-studied object, but also features a wealth of structures possibly related to (ongoing) planet formation.
In this paper we present new images of the HD100546 transition disk taken in PDI mode in the $H$, $K_{\rm s}$ and $L'$ filters. These data have a higher signal-to-noise than previous data sets allowing a more robust analysis of the disk morphology, and, in addition, in combination with a re-reduction of earlier data taken in 2006 \citep{quanz2011}, these data allow us to investigate possible changes in disk morphology and brightness over a baseline of $\sim$7 years.

\begin{deluxetable*}{cc@{\hspace{8pt}$\times$\hspace{8pt}}c@{\hspace{8pt}$\times$\hspace{8pt}}c@{\hspace{8pt}=\hspace{8pt}}cccccccc}
\centering
\tablecaption{Summary of observations. 
\label{table:observations}}           
\tablewidth{450px}
\tablehead{
\colhead{}    &  \multicolumn{4}{c}{Integration Time} & \colhead{} &  \multicolumn{4}{c}{Observing Conditions} \\ 
\cline{2-5} \cline{7-10} \\ 
\colhead{Filter} & \colhead{DIT\tablenotemark{a}}\hspace{12pt} & \colhead{NDIT\tablenotemark{a}}\hspace{16pt} & \colhead{NINT\tablenotemark{a}}\hspace{16pt} & \colhead{Total\tablenotemark{a}} & \colhead{} & \colhead{Airmass} & \colhead{Seeing\tablenotemark{b}} & \colhead{$\tau_0$\tablenotemark{c}} & \colhead{Coh. Energy\tablenotemark{d}}
}
\startdata
H & 0.3454s&80&27 (27) & 746s (746s)&&1.43&0.81$\arcsec$&1.9ms&41.7$\%$&\\K$_{\rm s}$ & 0.3454s&80&30 (28) & 829s (774s)&&1.50&0.86$\arcsec$&1.8ms&35.9$\%$&\\L & 0.175s&180&18 (16) & 567s (504s)&&1.63&1.03$\arcsec$&1.6ms&20.9$\%$&\vspace{3pt}\\H (cube mode) & 0.039s&1300 (975)&4 & 203s (152s)&&1.45&0.59$\arcsec$&2.6ms&51.2$\%$&\\K$_{\rm s}$ (cube mode) & 0.039s&2000 (1500)&3 & 234s (176s)&&1.44&0.70$\arcsec$&2.2ms&33.0$\%$&\vspace{3pt}\\H (2006) & 0.3454s&85&15 (15) & 440s (440s)&&1.58&1.09$\arcsec$&2.4ms&34.2$\%$&\\K$_{\rm s}$ (2006) & 0.3454s&85&13 (9) & 382s (264s)&&1.46&1.01$\arcsec$&2.7ms&40.6$\%$&\\
\enddata
\tablenotetext{a}{The detector integration time (DIT) multiplied by the number of integrations per frame (NDIT) multiplied by the number of integrations summed over all dither positions (NINT) gives the total integration time per retarder plate position. Numbers in brackets are the number of frames used and integration times achieved after frame selection was applied.}
\tablenotetext{b}{Average DIMM seeing in the optical during the observations, monitored by the seeing monitor at VLT.}
\tablenotetext{c}{Average coherence time of the atmosphere as calculated by the real time computer of the AO system.}
\tablenotetext{d}{Average coherent energy according to the ESO real time computer.}
\end{deluxetable*}

\section{Observations and data reduction}\label{observations_section}
The new observations were performed on the night of March 31, 2013 using the NAOS/CONICA (NACO) instrument mounted on UT4 (Yepun) of the Very Large Telescope (VLT) at Cerro Paranal, Chile, in the $H$, $K_{\rm s}$ and $L'$ filters. We used the SL27 camera (27 mas pixel$^{-1}$) in $HighDynamic$ mode ($HighWellDepth$ for the $L'$ filter) and read out in $Double RdRstRd$ mode ($Uncorr$ for the $L'$ filter). HD100546 is bright enough to saturate the detector in both the $H$ and $K_{\rm s}$ filter at the shortest detector integration time available in full frame mode (0.3454s). 
We used windowing in cube mode mode (only 256x256 of the 1024x1024 pixels of the NACO detector are read out, the shortest possible integration time is reduced to 0.039s) in order to get unsaturated images to study the innermost parts of the disk in the $H$ and $K_{\rm s}$ filter. These were also used to perform the photometric calibration as described in \citet{avenhaus2014}. There is a general uncertainty of $\sim$30$\%$ to this technique. In the $L'$ filter, the star was unsaturated at the shortest possible integration time of 0.175s and these data could be used for the photometric calibration directly.

In PDI mode, a Wollaston prism splits the incoming beam into an ordinary and extraordinary beam separated by 3.5$\arcsec$~on the detector. A polarimetric mask prevents the two beams from interfering, but limits the field of view to stripes of $\sim$27$\arcsec\times$3$\arcsec$. The rotatable half-wave retarder plate (HWP), controlling the orientation of the polarization was set to 0$^\circ$ / $-$45$^\circ$ to measure Stokes $Q$ and $-$22.5$^\circ$ / $-$67.5$^\circ$ to measure Stokes $U$. This means that we cycled through four retarder plate positions for each dither position and each integration. The total on-source integration times were 2984s, 3316s and 2268s in the $H$, $K_{\rm s}$ and $L'$ filter, respectively, and 811s / 936s in the $H$ and $K_{\rm s}$ filter using cube mode. Complementing these new data are the data taken on April 7, 2006, in the $H$ (1762s) and $K_{\rm s}$ (1527s) filter \citep[discussed in][]{quanz2011}, which we include in our analysis. A summary of the observations is given in Table \ref{table:observations}.

The data reduction procedure is described in detail in the appendix of \citet{avenhaus2014}. 
Two improvements to the pipeline are worth noting. First, we implemented a frame selection technique
to exclude frames that were taken when the adaptive optics (AO) performed poorly or are degraded in image quality for other reasons. We note the amount of frames selected and the resulting on-source integration time in Table \ref{table:observations}. Furthermore, for the $L'$ filter reduction, it was necessary to carefully subtract the high thermal background. To do this, from a given frame we subtracted another frame that was taken close in time, but at a different dither position.
For the cube mode images, each frame from the image stack was handled individually.

We then compute the images showing the tangential ($P_\perp$) and radial ($P_\parallel$) polarization directions, meaning polarization perpendicular to the line between the star and a given point in the image plane ($P_\perp$) and polarization parallel to this line ($P_\parallel$). We do this because single scattering off dust grains in a protoplanetary disk is expected to cause only polarization in the tangential direction, but not in the radial one. This technique has the advantage that $P_\perp$ gives an unbiased estimate of the polarized intensity $P$ \citep[c.f.][]{avenhaus2014}. However, strictly this is only true for disks that are either optically thin and where the signal is thus dominated by single scattering, or for optically thick disks seen face-on. In the case of inclined, optically thick disks, it only holds approximately. However, any deviation of the scattered light from being polarized in the tangential direction would show up in the $P_\parallel$ image and is thus included in our error estimates. The error from this effect (the polarization not being perfectly tangential) is significantly smaller than the other error sources in our images and can therefore be neglected. For comparison, we also use the conventional way of calculating $P$ ($P=\sqrt{Q^2+U^2}$).

\section{Results and Analysis}\label{results}
\label{sec:results}

\begin{figure*}
\centering
\includegraphics[width=0.95\textwidth]{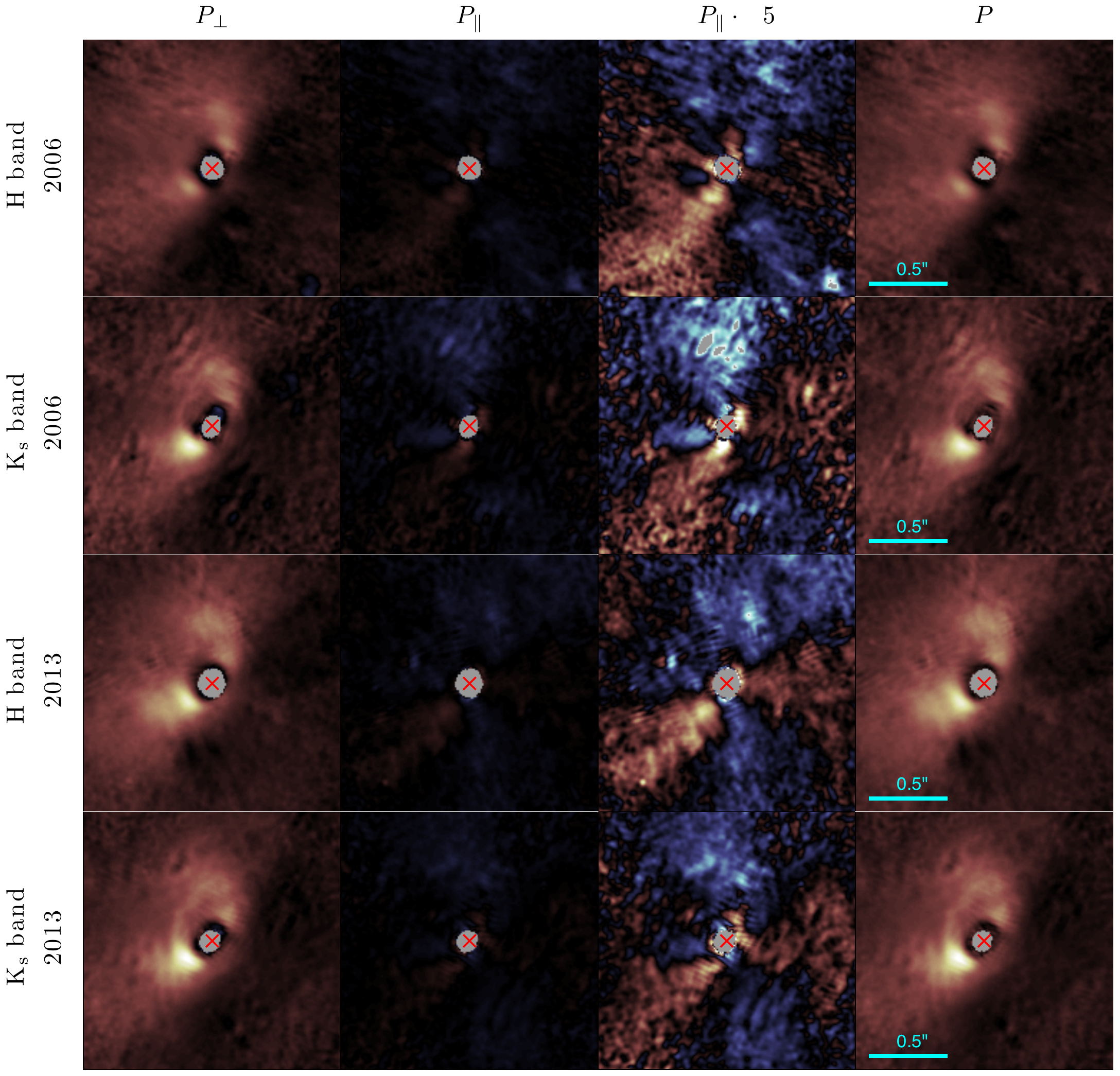}
\vspace{8pt}
\caption{NACO PDI results in $H$ and $K_{\rm s}$ filter from epochs 2006 and 2013. From left to right: $P_\perp$, capturing the structure of the disk, $P_\parallel$, which is expected to be zero and dominated by noise, $P_\parallel$ scaled by a factor of five to better show the noise signature, and $P$, which is identical to $P_\perp$ in the absence of any noise and when there is no rotation of the polarization due to multiple-scattering effects (see also text). Positive values are in orange, negative values in blue. The grey area in the center represents positions where no data is available due to saturation effects. The red cross marks the position of the star. North is up and east is to the left in all images. The images are 1.62\arcsec~($\sim$ 160 AU) on each side, they all show the same section of the disk. For reference, there is a scale in each of the $P$ images. All images scaled with $r^2$.
\label{images}}
\end{figure*}

\begin{figure*}
\centering
\includegraphics[width=0.95\textwidth]{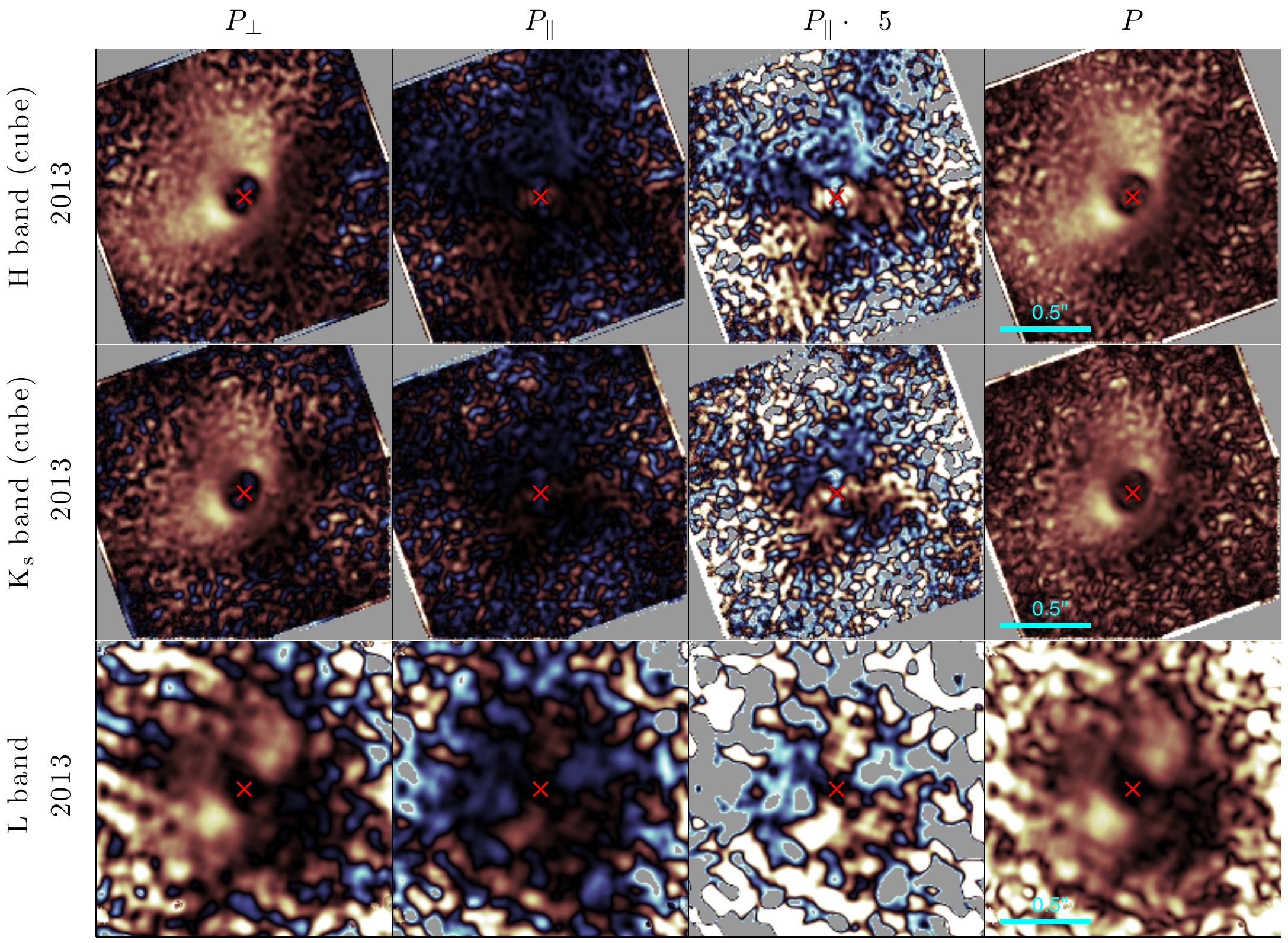}
\vspace{8pt}
\caption{Same as Figure \ref{images}, but for the $H$ and $K_{\rm s}$ cube-mode and the $L'$ filter observations.
\label{images2}}
\end{figure*}

The scattered light images of HD100546 in all filters and at both epochs are summarized in Figures \ref{images} and \ref{images2}. Besides $P_\perp$ (left), and $P$ (right), which are very similar to each other as expected, we also show $P_\parallel$ (as an indication of the noise level in the images) in two representations: Once scaled like the $P_\perp$ image (middle left) and once multiplied by a factor of 5 (middle right). We emphasize here that while these images only show the disk out to $\sim$0.8\arcsec, we can trace the disk to more than 1.5\arcsec~in every direction (see Figure \ref{fig:HD100546SB}).

The overall structure in the $P_\parallel$ images is similar in all $H$ and $K_{\rm s}$ observations. Static structure in $P_\parallel$ could hint towards a rotation in the polarization, but this is misleading: The structure seen here depends on the choice of reduction parameters, specifically on the inner and outer radius used for correcting the instrumental polarization \citep[see][]{avenhaus2014}. Because of this, we do not interpret the structure seen in these images, but note that because it is consistent in all datasets, the different final images can be compared very well relative to each other. 
The residuals in $P_\parallel$ are small compared to $P_\perp$. It can also be seen that the differences between $P$ and $P_\perp$ are small, but the $P$ images show slightly more noise very close to the star. 

The general structure of the disk is very well seen in all $H$ and $K_{\rm s}$ filter observations in both 2006 and 2013. The $L'$ filter observations suffer from more noise, but show similar structure in the regions where the SNR is high enough. The reason for the higher noise is the strong background emission in $L'$, which is orders of magnitude higher compared to the shorter wavelengths.

While the cube-mode and non-cube-mode observations in 2013 are comparable for the $H$ filter, the $K_{\rm s}$ filter cube mode observations appear darker (the observations were scaled to the same detector counts per time). The structure is similar. A possible explanation for this is that the signal is dampened by an effect similar to the one suppressing a polarization detection at very small separations (see discussion in Section \ref{sec:innerHole}), i.e. a smearing out of the butterfly pattern in the Stokes $Q$ and $U$ vectors due to the PSF of the observation. While the observing conditions were slightly better, the coherent energy was slightly worse (c.f. Table \ref{table:observations}).

\subsection{Global Scattering Signature}
\label{sec:globalScatteringSignature}

With our new data, we confirm the basic disk structure already described in \citep{quanz2011}: The major axis of the disk runs in southeast-northwest direction, and the brightest parts of the disk are roughly along this axis. The northeastern part of the disk appears brighter compared to the southwestern part. For the first time, we identify a dark lane between $\sim$0.2\arcsec~and $\sim$0.6\arcsec~on this forward-scattering side in all $H$ and $K_{\rm s}$ filter observations including the cube mode observations. The scattered light picks up (in this representation scaled with r$^2$) outside of $\sim$0.6\arcsec.

This dark lane together with the northeastern side of the disk appearing significantly brighter suggest that the grains in the disk are preferentially backscattering in polarization (scattering albedo multiplied with polarization fraction, which is what our data measure). Furthermore, the polarization efficiency in scattering usually peaks around 90$^\circ$ \citep[e.g.,][]{perrin2009}, which explains the two bright lobes in the southeast and northwest: The semi-major axis of the disk runs along this direction, and the scattering angle at these positions is close to 90$^\circ$ depending on the exact flaring angle.

\begin{figure}
\centering
\includegraphics[width=0.45\textwidth]{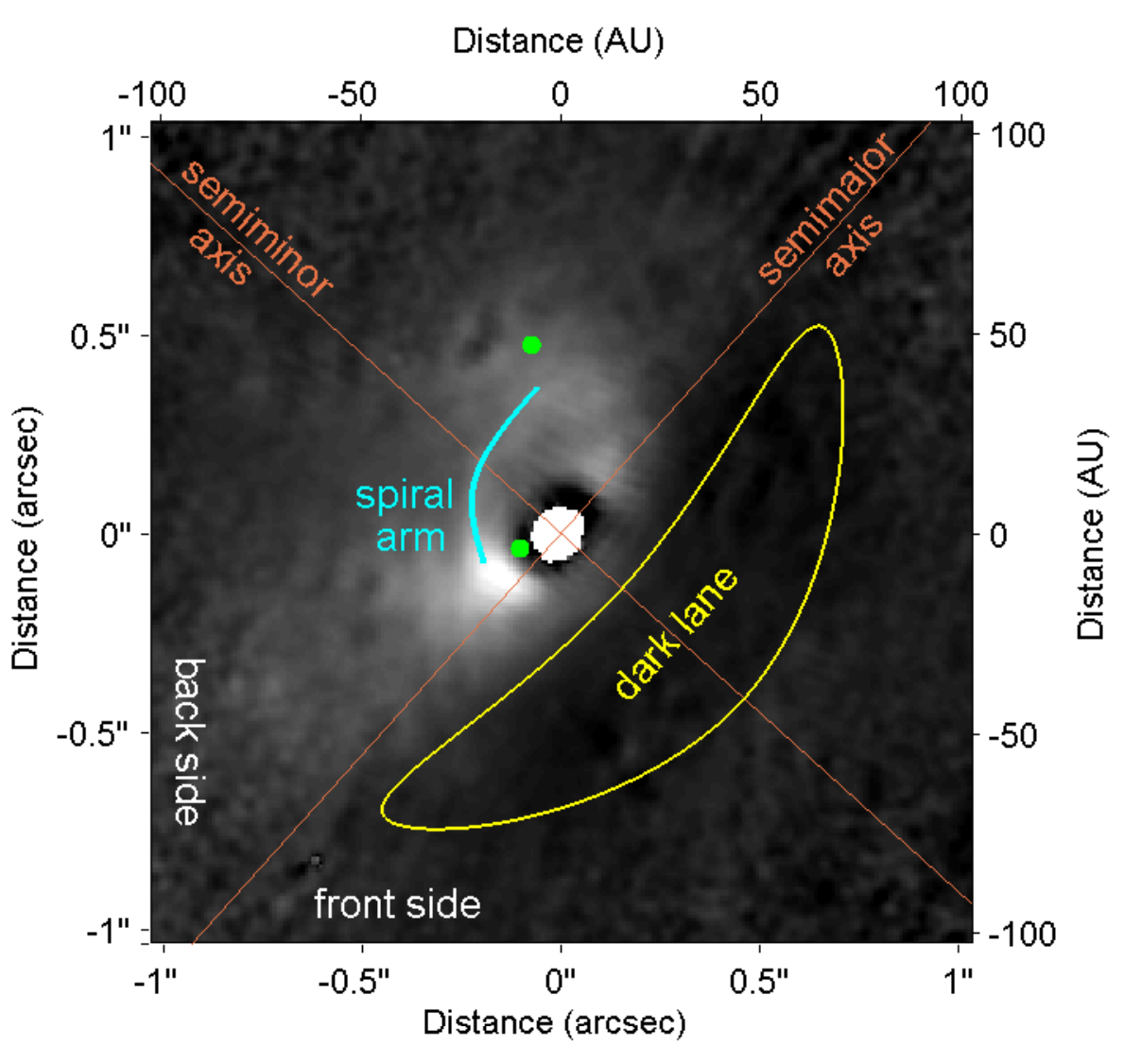}
\caption{\small Disk features seen in HD100546. The 2013 $K_{\rm s}$ filter data is overlaid with the main features detected in the disk. The dark lane in the southwest seems to fold around the stellar position. The spiral arm is marked in cyan, the positions of the two planet candidates from \citet{quanz2013a} and \citet{brittain2013} are marked in green.
\label{fig:diskStructure}}
\end{figure}

\begin{figure*}
\centering
\includegraphics[width=1\textwidth]{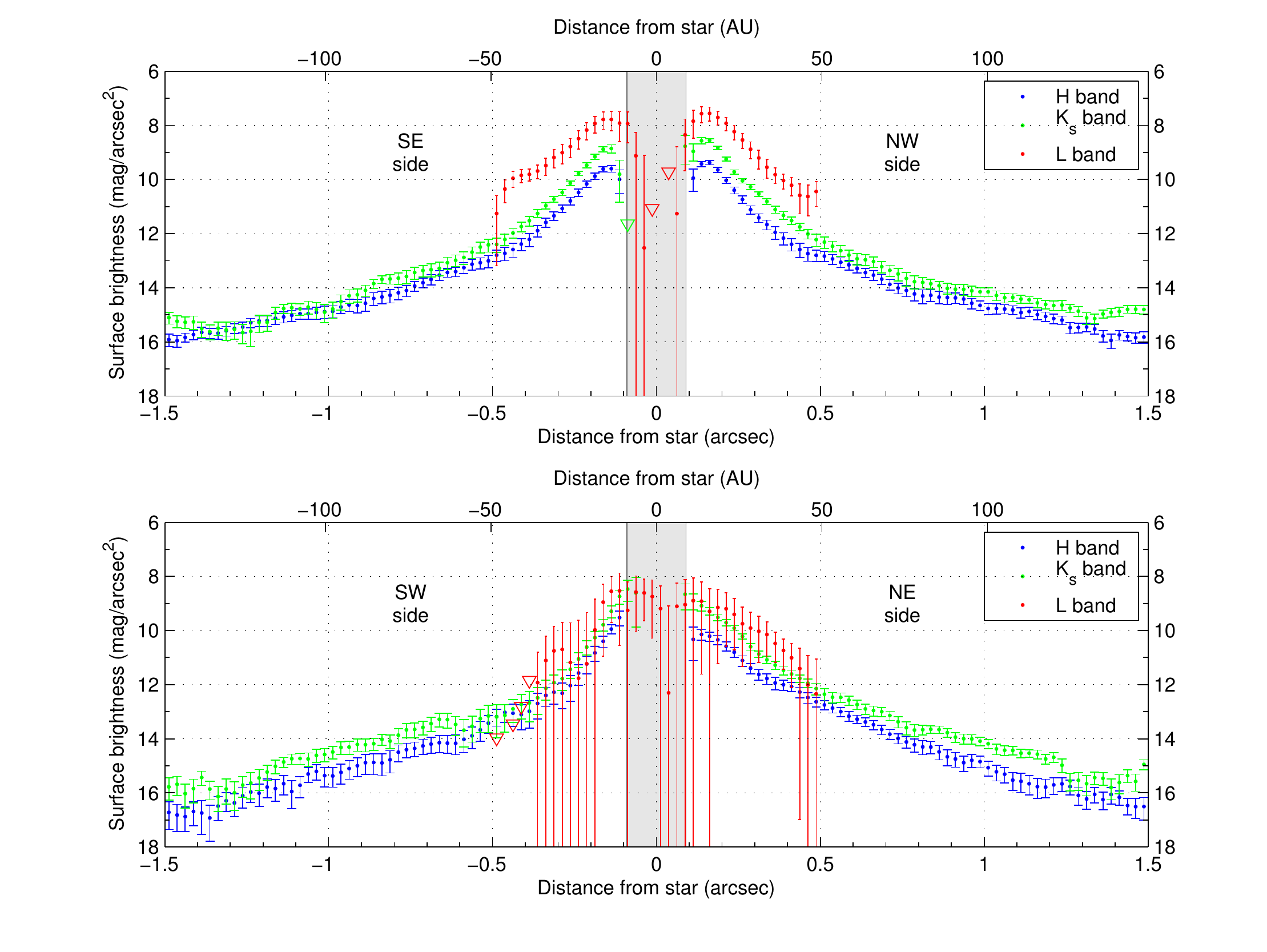}
\caption{\small Surface brightness plots of HD100546 for the $H$ (blue), $K_{\rm s}$ (black) and $L'$ (red) filter data of the 2013 saturated observations. The measurements were taken in 10$^\circ$ wedges along the semi-major (top) and semi-minor (bottom) axis of the disk. As the position angle for the semi-major axis, we use 138$^\circ$ \citep[][Section \ref{sec:diskPA} of this paper]{quanz2011}. The area not accessible with our data in the $H$ and $K_{\rm s}$ band is shaded in grey. The error bars represent 1$\sigma$ errors calculated in the same way as has been done for HD142527 \citep{avenhaus2014}. Downward-facing triangles represent 1$\sigma$ upper limits. The errors do not include a general calibration uncertainty of $\sim$30$\%$. They also do not account for the dampening effect of the PSF smearing described in Section \ref{sec:HD100546_PSF_effects}, which can be around one magnitude at the position of the inner rim. The $L'$ filter data is strongly dominated by noise outside $\sim$0.5\arcsec, which is why we restrict our plot to this distance. Along the semi-major axis, the inner hole is detected in all three filters and in both directions, while it is not seen in this representation along the semi-minor axis.
\label{fig:HD100546SB}}
\end{figure*}

The structures seen in the disk, along with the position of the two planet candidates \citep{quanz2013a, brittain2013}, are marked in Figure \ref{fig:diskStructure} on the left. With respect to the semi-minor axis, the dark lane in the southwest is relatively symmetric and seems to fold around the star. In the $H$ filter images, there seems to be a bridge of stronger scattering exactly in the direction of the semi-minor axis. This effect is weaker in the $K_{\rm s}$ filter images. The dark lane is not seen in the surface brightness plots (Figure \ref{fig:HD100546SB}) partly because of this and partly because these are not scaled with $r^2$.

While it is in principle possible that such a dark lane is produced by shadowing effects within the disk, i.e., a shadow cast from the inner rim, we deem this unlikely for two reasons. First, it is difficult to imagine that such a shadow appears on only one side of the disk and almost perfectly aligned with the semi-minor axis. Furthermore, the disk rotated by $\sim$75$^\circ$ at the position of the inner rim (c.f. Section \ref{sec:innerHole}) during our seven years baseline between the 2006 and 2013 observations, yet the dark lane stays at the same position.

Similar dark lanes have been seen for instance in non-polarimetric HST observations of IM Lupi \citep{pinte2008} and GM Aurigae \citep{schneider2003}. These authors explain the dark lane with a strongly inclined and flared disk, which causes a shadow on the forward-scattering side. Further out, where the disk becomes optically thin enough, the brightness increases again because scattering from the lower surface of the disk can be seen. It is questionable that this explanation works in the case of HD100546. First, the inclination and flaring angle derived by other authors for this disk \citep[$\sim$45-50$^\circ$ inclination, see discussion in Section \ref{sec:HD100546discussion}, and $\sim$7$^\circ$ flaring angle, see][]{benisty2010} are too small. Second, the disk would have to be optically thin in the near-IR at a radius of $\sim$100 AU. This, however, is in agreement with \citet{augereau2001}, who estimate the disk to be optically thin in the near-IR outside $\sim$80 AU.

A third possibility is that the dark lane results from the scattering function of the dust grains. This requires that the scattering angle varies across the disk in the direction of the semi-minor axis.
It also requires that the polarized scattering function has a minimum somewhere below 90$^\circ$ and increases again towards smaller scattering angles. This seems to be the case, as we discuss in more detail in Section \ref{sec:ScatteringFunction}. While we cannot explain the exact details of the polarized scattering curve (multiple scattering and dust grain properties both play a role here), we deem this explanation the most likely. 


\subsection{Surface Brightness Profiles}\label{sec:HD100546SB}

The surface brightness profiles in the $H$, $K_{\rm s}$ and $L'$ bands along the semi-major and semi-minor axes are shown in Figure \ref{fig:HD100546SB}. As can be seen, we are able to trace the disk significantly further than shown in the images in Figures \ref{images} and \ref{images2}, where we concentrate on the most interesting inner part of the disk. The inner region ($\lesssim$10 AU) contains no data for the $H$ and $K_{\rm s}$ band due to saturation and is marked in grey.

The numeric values of these surface brightnesses have to be treated with caution. As we discuss in Section \ref{sec:HD100546_PSF_effects}, the measured surface brightnesses are significantly dampened by the PDI technique. Because of this, we do not fit power laws to our data. However, we still observe that the slope of the surface brightness profiles is not constant. There seems to be a break between an inner region, where the slope is steeper, and an outer one, where the slope is less steep. The break can be observed at $\sim$40-50 AU in the semi-major axis and possibly a little further in in the semi-minor direction. This could be suggestive of changes in the dust grain properties \citep[see][]{pineda2014}.

Along the semi-major axis, the inner hole (see next section) is clearly detected. In the direction of the semi-major axis, the depletion at small radii is detected in all three filters, though it appears to be smaller in the $L'$ filter (this is likely due to stronger PSF smearing at longer wavelengths, see discussion in Section \ref{sec:HD100546_PSF_effects}). In the direction of the semi-minor axis, the gap is not seen so clearly in the surface brightness profiles. We note that these surface brightness profiles are generated from the saturated (not cube-mode) data. In the cube-mode images, the hole is clearly visible in both the semi-major and semi-minor direction.

\subsection{Disk Gap}\label{sec:innerHole}

Besides being visible in the surface brightness profiles, the disk gap can also be seen in all images. In the cube mode observations, the gap is detected very clearly. Taking into account the PSF smearing effect, we visually overlay the data with a ring for the inner rim in the various filters. The data in the $H$ and $K_{\rm s}$ filter (both normal and cube mode observations) are consistent with a circular inner rim at 14$\pm$2 AU and with a rather sharp inner rim edge which is only smeared out by the PSF. 

To analyze the degree of eccentricity of the inner cavity, we use the surface brightness images from the $H$ and $K_{\rm s}$ cube mode observations, estimated along the semi-major axis in wedges with 20$^\circ$ opening angle. We find the distance from the star in both directions along the semi-major axis where the surface brightness first reaches half the maximum brightness along this axis. We use the errors on the surface brightness, as estimated from the $P_\parallel$ images, to get an error estimate on this distance in both directions. Using these as Gaussian errors, we simulate 1'000'000 realizations of actual distances, taking into account the derived errors as well as the uncertainty of the position of the star. The star is unsaturated in the cube mode observations, and thus its position can be accurately determined. We estimate the uncertainty to be $\sim$5 mas ($\sim$0.5 AU / 0.2 pixels). From each pair of simulated values, we calculate the resulting eccentricity, allowing us to estimate probabilities for different values of the true eccentricity. We estimate from the $H$ filter results that the eccentricity is smaller than 0.113 with 95\% confidence and smaller than 0.178 with 99.8\% confidence. The $K_{\rm s}$ filter results lead to upper limits of 0.127 and 0.201 for these confidence levels, respectively. Combining the results from the two filters, we arrive at an upper limit of 0.085 at 95\% confidence and 0.133 at 99.8\% confidence. We conclude that the eccentricity along the semi-major axis is small, and our results are consistent with no eccentricity. However, we cannot make such statements for an eccentricity aligned with the semi-minor axis of the disk because of inclination effects.

The distance to the rim in the northeast seems to be larger than the one to the southwest, but this is consistent with an inclined, flared inner rim which is intrinsically circular around the star. Our data suggest that the inclination of this rim is below $\sim$50$^\circ$. The exact limit our data put on the inclination and the flaring is hard to determine, because the forward- and backward-scattering regions are intrinsically fainter. A high inclination would generate a more elliptic inner hole in the data, on the other hand the faintness in the direction of the semi-major axis reduces the optical visibility of such an ellipticity.

We do not detect any significant structure inside the disk cavity. The faint, ring-like structure seen in our cube-mode observations around the position of the star in the $P$ images is a noise artifact and not seen in the $P_\perp$ images. As discussed in Section \ref{sec:HD100546_PSF_effects}, the inner disk is not detectable with our observations due to PSF smearing effects if it resides at a radius of $\sim$3 AU or smaller.


\subsection{Inner Rim and Position Angle of the Disk}
\label{sec:diskPA}

The two bright points in the rim are at 127$^\circ$$\pm$5$^\circ$ / 126$^\circ$$\pm$6$^\circ$ ($H$ / $K_{\rm s}$ filter measurement for the bright peak in the southeast) and 333$^\circ$$\pm$7$^\circ$ / 327$^\circ$$\pm$6$^\circ$ ($H$ / $K_{\rm s}$ filter measurement for the fainter peak in the northwest) east of north. Combining these measurements, this means that they are 203$^\circ$$\pm$9$^\circ$ apart, i.e. not exactly opposite from each other, but slightly displaced w.r.t. the semi-major axis. The reason for this is most likely that the disk is not flat, but flared. This shifts the points of 90$^\circ$-scattering a little bit to the back side of the disk. While the measurements are not accurate enough to put constraints on either the inclination or the flaring angle, 
we can compare the two peaks individually to the adopted position angle of 138.0$^\circ$$\pm$3.9$^\circ$ from \citet{quanz2011}. The bright peak is displaced from this by 11$^\circ$$\pm$6$^\circ$ towards the back side of the disk, while the fainter peak is displaced by 12$^\circ$$\pm$6$^\circ$.

Turning this around, we can calculate the position angle (PA) of the disk by assuming that the two bright peaks are displaced from the semi-major axis by the same amount. This calculation yields a value of 138.2$^\circ$$\pm$3.0$^\circ$ when combining the data from $H$ and $K_{\rm s}$ filter, consistent with the adopted value. We stress at this point that our error estimate does not include systematic effects. The reflection points are intrinsically asymmetric in their brightness, which implies a physical difference between the two sides (southeast vs. northwest) of the disk. The southeast side of the inner rim is significantly brighter than the northwestern one in our 2013 observations in all three filters. This can be seen both in the images and in the surface brightness plots (Figure \ref{fig:HD100546SB}). However, we emphasize that these plots are along the semi-major axis, which does not pass through the brightest areas exactly. We find the peak in the southeast to be brighter than the peak in the northwest by a factor of $1.67\pm0.33$, $1.92\pm0.33$ and $1.51\pm0.70$ in the $H$, $K_{\rm s}$ and $L'$ filter, respectively. The errors on these values have been estimated from the residuals in the $P_\parallel$ images. We exclude the possibility that this difference in brightness is caused by one side of the disk being closer to the star, because the asymmetry is too strong to be explained by such an effect and the distance of both bright spots to the star is similar. In our 2006 observations, the asymmetry is only seen in the $K_{\rm s}$ filter. Also, the $H$ filter shows a significantly weaker overall scattering signal.

\subsection{Spiral Arm}

A new feature detected with our data is a faint spiral arm extending from the bright southeastern region in a clockwise manner towards the north and then the west. This feature is most clearly seen in the 2013 $K_{\rm s}$ filter data, but can also be spotted in the 2006 $K_{\rm s}$ and 2013 $H$ filter data. We are confident that this feature is not an artifact from the data reduction because it can be seen in several datasets. 

\subsection{Scattering function}\label{sec:ScatteringFunction}

\begin{figure*}
\centering
\includegraphics[width=0.95\textwidth]{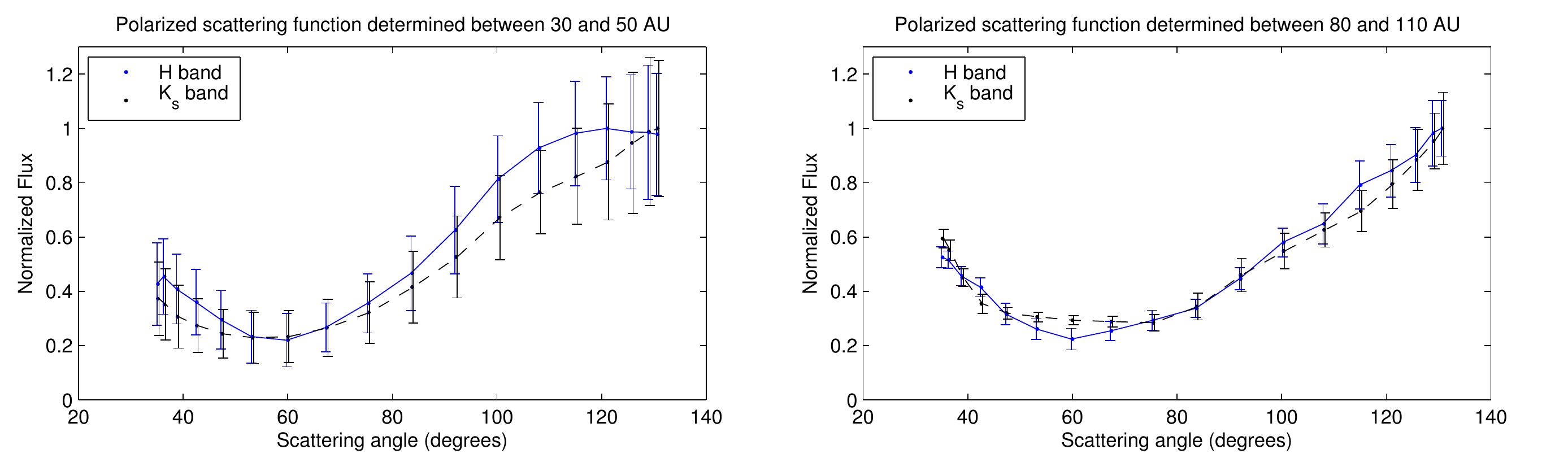}
\caption{\small Scattering function determined from our data between 30 and 50 AU (left) and 80 and 110 AU (right). We use an inclination of 48$^\circ$ and a constant flaring angle of 7$^\circ$ for these calculations, the same values used by \citet{quanz2011}. The values have been normalized to the maximum value for each graph. For discussion, see text.\label{fig:scatterFunction}
}
\end{figure*}

HD100546 is one of the few disks which is suitably inclined to determine the scattering function over a large range of scattering angles, and we derive it between $\sim$35$^\circ$ and $\sim$130$^\circ$. In the case of PDI data, one measures the product of the scattering albedo of the dust grains and their polarization efficiency. To calculate the scattering function, we assume the flaring angle of the disk to be constant at 7$^\circ$ \citep[c.f.][]{benisty2010} and an inclination of 48$^\circ$, the same values used by \citet{quanz2011}. We calculate the scattering angles at two different annuli (30 to 50 and 80 to 110 AU) and show our results in Figure \ref{fig:scatterFunction}. 

As can be seen, the grains are preferentially backscattering in polarized light. The scattering function reaches a minimum at $\sim$60$^\circ$ and rises again towards smaller values. A forward-scattering peak could explain the brighter bridge of light towards the southwest described in Section \ref{sec:globalScatteringSignature}. The values between the $H$ and $K_{\rm s}$ filter are consistent at both annuli, but seem to differ between the two. For the outer annulus, the curve seems to be overall flatter, rising more strongly towards small scattering angles ($\lesssim$50$^\circ$) and rising later towards larger scattering angles ($\gtrsim$80$^\circ$). An explanation for this behavior could be that the grain properties vary with radius. Another possibility is that the flaring angle is not constant with radius, but increases towards the outer regions of the disk. In this case, the analysis at 80 to 110 AU would probe smaller scattering angles, moving the entire graph to the left by a few degrees - and making the two curves more consistent with each other. This, as well as the strong scattering at small scattering angles, is in agreement with the interpretation of the dark lane in Section \ref{sec:globalScatteringSignature}. Because of that, we prefer this second interpretation, without being able to exclude the possibility of grain properties varying with disk radius.

\subsection{Disk Color}

The 2013 $H$ and $K_{\rm s}$ filter images of the disk and surface brightness plots show no color variations which we would deem significant. In the $L'$ filter, the difference between the semi-major and semi-minor axis seems stronger compared to the $H$ and $K_{\rm s}$ filter, but the SNR is very low in the semi-minor direction.

The three filters allow us to determine the overall scattering color of the disk. To do this, and to be able to also determine the scattered-light flux in the low-SNR $L'$ filter data, we calculate the total scattered light in an annulus between 0.12\arcsec~and 0.3\arcsec. By comparison to the stellar flux, we can then determine the scattering color of the disk in this annulus. Because we do not need to convert to 2MASS magnitudes in between, this direct comparison yields color estimates with smaller errors. The resulting colors are 0.19$\pm$0.11 mag in [$H$]-[$K_{\rm s}$], -1.08$\pm$0.35 mag in [$H$]-[$L'$] and -1.27$\pm$0.35 mag in [$K_{\rm s}$]-[$L'$], meaning that the disk scattering is weaker in the $L'$ filter. Between the $H$ and $K_{\rm s}$ filter, the color is almost grey, consistent with being zero. We emphasize at this point that the PSF smearing effect discussed in the next section can have an influence on color. Specifically, it could dampen the longer wavelengths stronger, which would particularly affect the $L'$ filter measurements. 
We estimate that this effect could explain only part of the lack in the $L'$ filter, though, and that the scattering in the $L'$ filter is truly significantly weaker than the scattering in $H$ and $K_{\rm s}$ filter by at least half a magnitude.

\subsection{PSF smearing effects}
\label{sec:HD100546_PSF_effects}

\begin{figure*}
\centering
\plotone{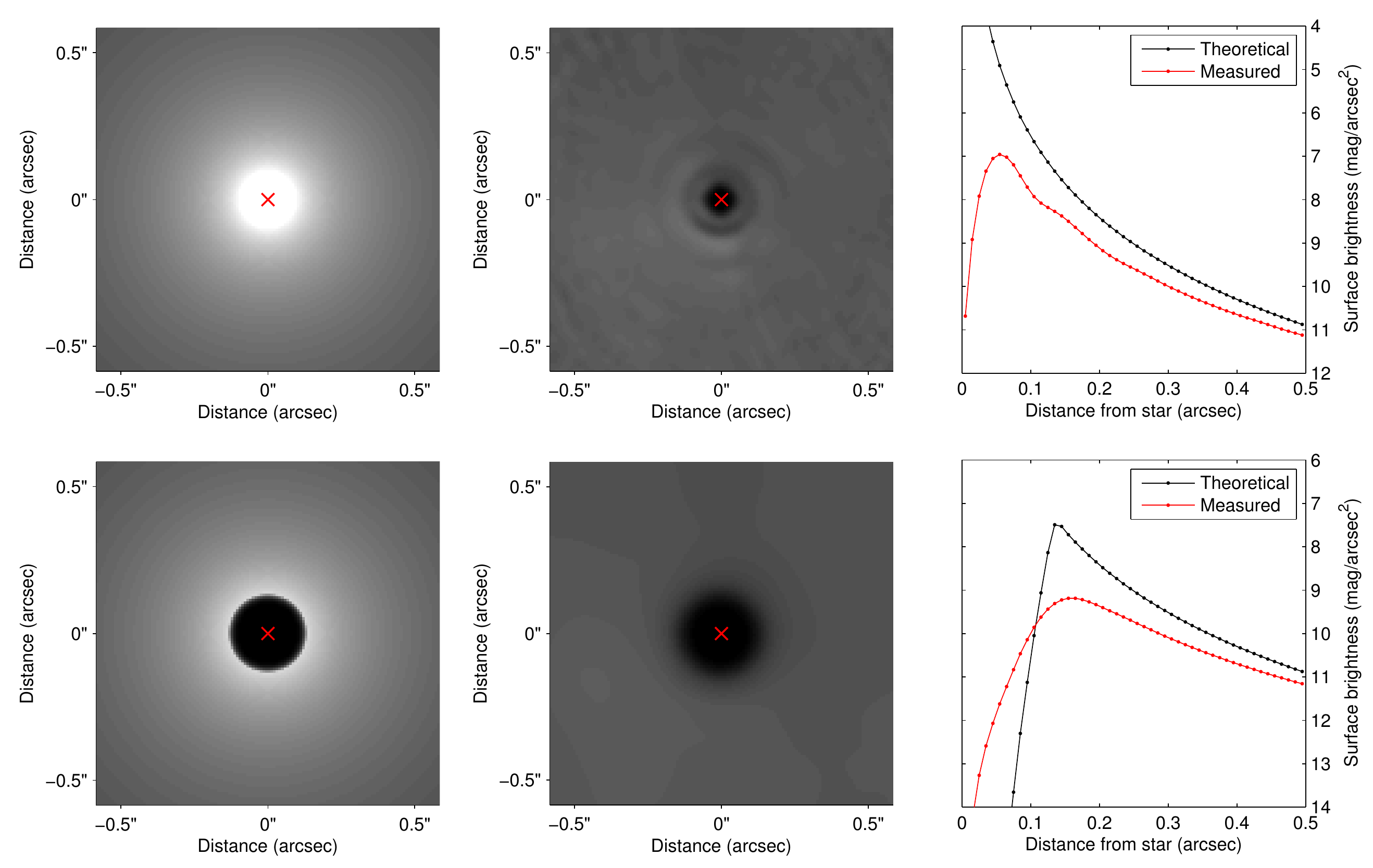}
\caption{\small Two theoretical disk models at infinite resolution and surface brightness falling off as $r^{-2.5}$ compared with the same disk models (Stokes $Q$ and $U$ vectors separately) convolved with the PSF taken from unsaturated observations ($K_{\rm s}$ filter cube mode). Left: Model. Middle: Expected observations. Right: Surface brightness of model and expected observations compared. As can be seen, a hole appears at the stellar position as an artifact, together with structures in the disk which stem from the diffraction rings of the PSF. The polarimetric signal is also significantly dampened. All images scaled in the same way.\label{fig:PolDampingCloseToStar}}
\end{figure*}

PSF smearing affects all observations of protoplanetary disks, but in the case of polarimetric differential imaging, these effects are more complicated because the polarized flux is not measured directly, but derived from the Stokes vectors $Q$ and $U$. In contrast to direct flux measurements, both the $Q$ and the $U$ vector can be negative and usually show a butterfly pattern for protoplanetary disks, as shown for instance in \citet{quanz2011}.

As a consequence, the polarization signal ($P$ and $P_\perp$, respectively) calculated from these vectors is not only smeared, but also dampened close to the star. The butterfly pattern in the Stokes $Q$ and $U$ vector at the center of the image (close to the star) is washed out by the PSF, and as a result, even disks without an inner hole (or an unresolvable hole at 0.001\arcsec) would show a hole in the polarized light as an artifact. This is shown in Figure \ref{fig:PolDampingCloseToStar}. Furthermore, the typical dilution of local features is seen, smearing out the scattered-light signal from the inner rim of the disk (lower part of Figure \ref{fig:PolDampingCloseToStar}). The effects can be severe, in this example dampening the flux from the inner rim by more than one magnitude. The effect is stronger at longer wavelengths, where the PSF is larger.

We emphasize at this point that our calculations show that this effect is clearly not the reason for the inner hole in our observations. Its size would be smaller and also depend on the observing conditions and wavelength, with the hole being larger at longer wavelengths, which not what we see. However, this effect erases the signal very close to the star (inside of $\sim$0.03$\arcsec$). Thus, an inner disk at $\sim$3 AU or less would not be detected in our observations. 

\subsection{Comparison of 2006 and 2013 Data}\label{sec:HD100546comp}

Comparing the results from the 2006 data discussed in \citet{quanz2011} to our new results, we have to refine some of the findings. In that paper, a hole was seen towards the north, in the direction where a protoplanet candidate was subsequently detected in \citet{quanz2013a}. The damping of the Stokes $U$ vector described in \citet{avenhaus2014} was not taken into account in that earlier work, because it was only realized later when more observations were available. Applying the corrections calculated from the data as described in the appendix of \citet{avenhaus2014}, the hole mostly disappears. There still is a slight depletion in polarized scattered light at the inner rim of the disk, but the interpretation of a large-scale hole in polarized scattered light in the northern direction is not supported by our results seen in Figures \ref{images} and \ref{images2}.

The detection of the inner hole at $sim$14 AU, though, is clearly supported. Also, a clump seen in the north-northwestern direction of the disk still seems to be present, at least, the disk is not smooth in this direction. The position angle determined in \citet{quanz2011} is consistent with the position angle we calculate from the 2013 data using a different technique.

The dataset which differs most from the rest is the 2006 $H$ filter data. We do not have a clear explanation for this and cannot completely exclude the possibility that instrumental and data reduction effects cause this.

We would deem the 2013 data more reliable, because it has longer integration times and thus better SNR. In addition the data were taken with a better understanding of the instrument (improved setup). We also checked our calibration w.r.t. Stokes $Q$ and $U$ by rotating the field by 45$^\circ$ in the middle of the $H$ filter observations. The results from this test clearly support our interpretations and calculations found in the Appendix of \citet{avenhaus2014}.


\section{Discussion}\label{sec:HD100546discussion}

Measurements of the PA of the disk, which we estimate at 138.2$^\circ$$\pm$3.0$^\circ$, vary strongly and are often not consistent with each other. Measurements found in the literature range from 127$^\circ$$\pm$5$^\circ$ \citep{grady2001, pantin2000} through 145$^\circ$$\pm$5$^\circ$ \citep{ardila2007, panic2012} up to 161$^\circ$$\pm$5$^\circ$ \citep{augereau2001}. 
Our data only allow us to constrain the inclination of the disk to smaller than $\sim$50$^\circ$. This is consistent with literature values of 42$^\circ$$\pm$5$^\circ$ \citep{ardila2007}, 50$^\circ$$\pm$5$^\circ$ \citep{pantin2000}, 53$^\circ$$\pm$8$^\circ$ \citep{panic2012}, 49$^\circ$$\pm$4$^\circ$ \citet{grady2001}, 45$^\circ$$\pm$15$^\circ$ \citet{liu2003}, and 51$^\circ$$\pm$3$^\circ$ \citet{augereau2001}. \citet{benisty2010} found a scale height for the surface layer of 12AU at a distance of 100AU, which is equivalent to a flaring angle of 7$^\circ$. Our results suggest that the flaring angle varies with radius, both because of the dark lane found in the disk and because of the scattering function which we derive at two different radii.


The radius of the inner rim has been estimated from observations in the MIR to be $\sim$13 AU \citep{panic2012, tatulli2011}. \citet{vanderplas2009} and \citet{brittain2009} see a peak in the ro-vibrational CO line emission also at $\sim$13 AU. This is consistent with our measurement from the $H$ and $K_{\rm s}$ filter data of 14$\pm$2 AU for the radius of the inner rim. However, we do not see a gradual increase in scattered light between $\sim$10 AU and $\sim$25 AU as suggested by \citet{mulders2013b} for the MIR emission. Interestingly, \citet{liskowsky2012} suggest an eccentric inner rim of the disk along the semi-major axis with an eccentricity of $0.18^{+0.12}_{-0.11}$. We cannot confirm any eccentricity, from our scattered-light NIR data. As discussed, we exclude eccentricities greater than 0.133 at 99.8$\%$ confidence.

The time variable properties of the CO ro-vibrational emission from HD100546 led \citet{brittain2013} to infer the presence of a source of excess CO ro-vibrational emission that orbits the star at a distance of ~13 AU. The position and velocity information obtained through CO spectroastrometry located the excess source at a PA of $-2\pm10$ degrees in 2006 and $102\pm10$ degrees in 2013. That is, in 2013, the inner companion would be located close to the position of the SE peak, and in 2006, the companion would have been located far away. This is suggestive of a connection between the brightness asymmetry seen in our images and the companion, which orbits very close to the inner rim of the disk. If the companion was able to stir up the inner rim, increasing its scale-height, this could naturally explain the brighter scattering. 
While this is a tempting explanation and works well in the $H$ filter, where the asymmetry is much weaker in the 2006 observations, it does not for the $K_{\rm s}$ filter data, where the asymmetry is also present in 2006, when the companion was far away. This emphasizes the need for multi-color observations, but also makes an interpretation of this asymmetry challenging.

Another possible interpretation for the differences between the 2006 and 2013 measurements would be that either the illumination or the structure of the disk changed. 
The orbital timescale at the position of the inner rim ($\sim$14 AU) is $\sim$34 yr 
and $\sim$0.23 yr at 0.5 AU. Changes in the inner disk casting a shadow onto the inner rim of the outer disk could thus be responsible for the detected variations because they happen on timescales faster than our $\sim$7 yr baseline. However, we would expect them to influence both the $H$ and $K_{\rm s}$ filter. A change in the grain properties would also be a possibility, but we would do not expect grain properties to change significantly on such short timescales, unless there is an inherent asymmetry in the azimuthal direction which would rotate by $\sim$75$^\circ$ between our observations.

We are not able to detect the inner disk. This is not surprising given the fact that we are unable to detect even strong scattering at radii smaller than $\sim$3 AU due to the PSF smearing effects discussed in Section \ref{sec:HD100546_PSF_effects}. The inner disk is less than 0.7 AU and likely even less than 0.3 AU in size \citep{panic2012, mulders2013b}. Even if the disk extends out to $\sim$4 AU as suggested by \citet{tatulli2011}, it is unclear whether we would detect it.


The newly detected spiral arm 
has a direction consistent with spiral arms seen further out in the disk. The spiral arm has no obvious connection to any of these spiral arms detected by either \citet{ardila2007} or \citet{boccaletti2013}. It is important to remember that the scattered light traces the surface rather than the mid-plane of the disk, so we do not know whether this spiral arm represents a surface density enhancement or just a feature on the disk surface. 
ALMA observations tracing the mid-plane of the disk with comparable spatial resolution might be able to help answer this question.

The two companions suggested to orbit in this disk \citep{quanz2013a, brittain2013} should have an impact on the disk. While the inner companion seems to be responsible for the gap in the disk, and could be related to the brightness asymmetry of the inner rim, we see no obvious effect of the outer companion. The disk at this position seems to be relatively smooth. We do not see any evidence for a disk gap formed by the planet, alhtough a sufficiently small gap would not be detected with our observations. The gap would have to be significantly smaller than the our spatial resolution, though. A causal connection to the spiral arm is possible, but unclear. A connection to the break in the surface brightness profile around 0.5$\arcsec$~is also conceivable, but speculative.



\section{Conclusion}\label{conclusion}

The data presented in this paper clearly resolve the circumstellar environment of HD100546 close to the star at high SNR. 
The inner hole is detected with a radius of 14$\pm$2 AU and an inclination of less than $\sim$50$^\circ$. Some of the other disk features are puzzling. The general structure of the disk is well explained by preferentially backscattering grains, making the northeastern side of the disk the far and the southwestern side of the disk the near side. This also gives a natural explanation for the bright spots at the inner rim along the semi-major axis due to the scattering angle of $\sim$90$^\circ$. As a side effect, these scattering peaks allow us to constrain the position angle of the semi-major axis to 138.2$^\circ$$\pm$3.0$^\circ$. We emphasize that the error given here does not include possible systematic errors which could arise from intrinsic differences between the northwestern and southeastern side of the inner rim, but we expect such an error to be small (on the order of the statistical error or smaller). The disk hole towards the north detected by \citet{quanz2011} is not confirmed with our new data. It seems to be an artifact of the diminished flux in the Stokes $U$ vector, which we could correct for in this paper.

The dark lane in the near side of the disk is likely an effect of the polarized scattering function of the grains. The scattering function has a broad minimum at $\sim$60$^\circ$. This, together with the differences in the scattering function derived at two different radii, supports the interpretation of a flaring angle increasing with radius in the disk. To understand the dust scattering in detail, we would need a complete, self-consistent radiative transfer model of the disk, from which artificial PDI images could be produced. Such an analysis is beyond the scope of this paper and is left for future investigations.

Another unexplained penomenon is the brightness asymmetry of the disk rim, with the southeast side being significantly brighter in the $H$, $K_{\rm s}$ and $L'$ filter. The asymmetry cannot be caused by an ellipticity of the inner rim, but could be related to the companion orbiting within this rim. This connection, however, is speculative. It would be consistent with the fact that the inner companion should be close to the bright spot in early 2013 \citep{brittain2013}, but the asymmetry has been detected in the $K_{\rm s}$ band in 2006 as well.

The newly detected spiral arm could also have its origin in the companion(s), but again, this is unclear. We do not find a connection with any of the spiral arms detected by other authors. ALMA observations at similar resolution would help answer the questions about the nature of this feature, and would also allow us to determine whether the spiral arm is a surface feature of the disk or whether it is also present in the surface density.

\acknowledgments
This research has made use of the SIMBAD database, operated at CDS, Strasbourg, France. We thank the staff at VLT for their excellent support during the observations. This work is supported by the Swiss National Science Foundation.

Basic research in infrared astronomy at the Naval Research Laboratory is supported by 6.1 base funding.

S.D.B.  acknowledges support for this work from the National Science Foundation under grant number  AST-0954811.



{\it Facilities:} \facility{VLT:Yepun (NACO)}

\bibliographystyle{apj.bst}
\bibliography{HD100546.bib}

\begin{thebibliography}{41}
\expandafter\ifx\csname natexlab\endcsname\relax\def\natexlab#1{#1}\fi

\bibitem[{{Acke} \& {van den Ancker}(2006)}]{acke2006}
{Acke}, B. \& {van den Ancker}, M.~E. 2006, \aap, 449, 267

\bibitem[{{Ardila} {et~al.}(2007){Ardila}, {Golimowski}, {Krist}, {Clampin},
  {Ford}, \& {Illingworth}}]{ardila2007}
{Ardila}, D.~R., {Golimowski}, D.~A., {Krist}, J.~E., {Clampin}, M., {Ford},
  H.~C., \& {Illingworth}, G.~D. 2007, \apj, 665, 512

\bibitem[{{Augereau} {et~al.}(2001){Augereau}, {Lagrange}, {Mouillet}, \&
  {M{\'e}nard}}]{augereau2001}
{Augereau}, J.~C., {Lagrange}, A.~M., {Mouillet}, D., \& {M{\'e}nard}, F. 2001,
  \aap, 365, 78

\bibitem[{{Avenhaus} {et~al.}(2014){Avenhaus}, {Quanz}, {Schmid}, {Meyer},
  {Garufi}, {Wolf}, \& {Dominik}}]{avenhaus2014}
{Avenhaus}, H., {Quanz}, S.~P., {Schmid}, H.~M., {Meyer}, M.~R., {Garufi}, A.,
  {Wolf}, S., \& {Dominik}, C. 2014, \apj, 781, 87

\bibitem[{{Ayliffe} \& {Bate}(2009)}]{ayliffe2009}
{Ayliffe}, B.~A. \& {Bate}, M.~R. 2009, \mnras, 397, 657

\bibitem[{{Benisty} {et~al.}(2010){Benisty}, {Tatulli}, {M{\'e}nard}, \&
  {Swain}}]{benisty2010}
{Benisty}, M., {Tatulli}, E., {M{\'e}nard}, F., \& {Swain}, M.~R. 2010, \aap,
  511, A75

\bibitem[{{Boccaletti} {et~al.}(2013){Boccaletti}, {Pantin}, {Lagrange},
  {Augereau}, {Meheut}, \& {Quanz}}]{boccaletti2013}
{Boccaletti}, A., {Pantin}, E., {Lagrange}, A.-M., {Augereau}, J.-C., {Meheut},
  H., \& {Quanz}, S.~P. 2013, \aap, 560, A20

\bibitem[{{Bouwman} {et~al.}(2003){Bouwman}, {de Koter}, {Dominik}, \&
  {Waters}}]{bouwman2003}
{Bouwman}, J., {de Koter}, A., {Dominik}, C., \& {Waters}, L.~B.~F.~M. 2003,
  \aap, 401, 577

\bibitem[{{Brittain} {et~al.}(2009){Brittain}, {Najita}, \&
  {Carr}}]{brittain2009}
{Brittain}, S.~D., {Najita}, J.~R., \& {Carr}, J.~S. 2009, \apj, 702, 85

\bibitem[{{Brittain} {et~al.}(2013){Brittain}, {Najita}, {Carr}, {Liskowsky},
  {Troutman}, \& {Doppmann}}]{brittain2013}
{Brittain}, S.~D., {Najita}, J.~R., {Carr}, J.~S., {Liskowsky}, J., {Troutman},
  M.~R., \& {Doppmann}, G.~W. 2013, \apj, 767, 159

\bibitem[{{Cutri} {et~al.}(2003){Cutri}, {Skrutskie}, {van Dyk}, {Beichman},
  {Carpenter}, {Chester}, {Cambresy}, {Evans}, {Fowler}, {Gizis}, {Howard},
  {Huchra}, {Jarrett}, {Kopan}, {Kirkpatrick}, {Light}, {Marsh}, {McCallon},
  {Schneider}, {Stiening}, {Sykes}, {Weinberg}, {Wheaton}, {Wheelock}, \&
  {Zacarias}}]{cutri2003}
{Cutri}, R.~M., {Skrutskie}, M.~F., {van Dyk}, S., {Beichman}, C.~A.,
  {Carpenter}, J.~M., {Chester}, T., {Cambresy}, L., {Evans}, T., {Fowler}, J.,
  {Gizis}, J., {Howard}, E., {Huchra}, J., {Jarrett}, T., {Kopan}, E.~L.,
  {Kirkpatrick}, J.~D., {Light}, R.~M., {Marsh}, K.~A., {McCallon}, H.,
  {Schneider}, S., {Stiening}, R., {Sykes}, M., {Weinberg}, M., {Wheaton},
  W.~A., {Wheelock}, S., \& {Zacarias}, N. 2003, {2MASS All Sky Catalog of
  point sources.}

\bibitem[{{de Winter} {et~al.}(2001){de Winter}, {van den Ancker}, {Maira},
  {Th{\'e}}, {Djie}, {Redondo}, {Eiroa}, \& {Molster}}]{dewinter2001}
{de Winter}, D., {van den Ancker}, M.~E., {Maira}, A., {Th{\'e}}, P.~S.,
  {Djie}, H.~R.~E.~T.~A., {Redondo}, I., {Eiroa}, C., \& {Molster}, F.~J. 2001,
  \aap, 380, 609

\bibitem[{{Garufi} {et~al.}(2013){Garufi}, {Quanz}, {Avenhaus}, {Buenzli},
  {Dominik}, {Meru}, {Meyer}, {Pinilla}, {Schmid}, \& {Wolf}}]{garufi2013}
{Garufi}, A., {Quanz}, S.~P., {Avenhaus}, H., {Buenzli}, E., {Dominik}, C.,
  {Meru}, F., {Meyer}, M.~R., {Pinilla}, P., {Schmid}, H.~M., \& {Wolf}, S.
  2013, \aap, 560, A105

\bibitem[{{Grady} {et~al.}(2013){Grady}, {Muto}, {Hashimoto}, {Fukagawa},
  {Currie}, {Biller}, {Thalmann}, {Sitko}, {Russell}, {Wisniewski}, {Dong},
  {Kwon}, {Sai}, {Hornbeck}, {Schneider}, {Hines}, {Moro Mart{\'{\i}}n},
  {Feldt}, {Henning}, {Pott}, {Bonnefoy}, {Bouwman}, {Lacour}, {Mueller},
  {Juh{\'a}sz}, {Crida}, {Chauvin}, {Andrews}, {Wilner}, {Kraus}, {Dahm},
  {Robitaille}, {Jang-Condell}, {Abe}, {Akiyama}, {Brandner}, {Brandt},
  {Carson}, {Egner}, {Follette}, {Goto}, {Guyon}, {Hayano}, {Hayashi},
  {Hayashi}, {Hodapp}, {Ishii}, {Iye}, {Janson}, {Kandori}, {Knapp}, {Kudo},
  {Kusakabe}, {Kuzuhara}, {Mayama}, {McElwain}, {Matsuo}, {Miyama}, {Morino},
  {Nishimura}, {Pyo}, {Serabyn}, {Suto}, {Suzuki}, {Takami}, {Takato},
  {Terada}, {Tomono}, {Turner}, {Watanabe}, {Yamada}, {Takami}, {Usuda}, \&
  {Tamura}}]{grady2013}
{Grady}, C.~A., {Muto}, T., {Hashimoto}, J., {Fukagawa}, M., {Currie}, T.,
  {Biller}, B., {Thalmann}, C., {Sitko}, M.~L., {Russell}, R., {Wisniewski},
  J., {Dong}, R., {Kwon}, J., {Sai}, S., {Hornbeck}, J., {Schneider}, G.,
  {Hines}, D., {Moro Mart{\'{\i}}n}, A., {Feldt}, M., {Henning}, T., {Pott},
  J.-U., {Bonnefoy}, M., {Bouwman}, J., {Lacour}, S., {Mueller}, A.,
  {Juh{\'a}sz}, A., {Crida}, A., {Chauvin}, G., {Andrews}, S., {Wilner}, D.,
  {Kraus}, A., {Dahm}, S., {Robitaille}, T., {Jang-Condell}, H., {Abe}, L.,
  {Akiyama}, E., {Brandner}, W., {Brandt}, T., {Carson}, J., {Egner}, S.,
  {Follette}, K.~B., {Goto}, M., {Guyon}, O., {Hayano}, Y., {Hayashi}, M.,
  {Hayashi}, S., {Hodapp}, K., {Ishii}, M., {Iye}, M., {Janson}, M., {Kandori},
  R., {Knapp}, G., {Kudo}, T., {Kusakabe}, N., {Kuzuhara}, M., {Mayama}, S.,
  {McElwain}, M., {Matsuo}, T., {Miyama}, S., {Morino}, J.-I., {Nishimura}, T.,
  {Pyo}, T.-S., {Serabyn}, G., {Suto}, H., {Suzuki}, R., {Takami}, M.,
  {Takato}, N., {Terada}, H., {Tomono}, D., {Turner}, E., {Watanabe}, M.,
  {Yamada}, T., {Takami}, H., {Usuda}, T., \& {Tamura}, M. 2013, \apj, 762, 48

\bibitem[{{Grady} {et~al.}(2001){Grady}, {Polomski}, {Henning}, {Stecklum},
  {Woodgate}, {Telesco}, {Pi{\~n}a}, {Gull}, {Boggess}, {Bowers}, {Bruhweiler},
  {Clampin}, {Danks}, {Green}, {Heap}, {Hutchings}, {Jenkins}, {Joseph},
  {Kaiser}, {Kimble}, {Kraemer}, {Lindler}, {Linsky}, {Maran}, {Moos}, {Plait},
  {Roesler}, {Timothy}, \& {Weistrop}}]{grady2001}
{Grady}, C.~A., {Polomski}, E.~F., {Henning}, T., {Stecklum}, B., {Woodgate},
  B.~E., {Telesco}, C.~M., {Pi{\~n}a}, R.~K., {Gull}, T.~R., {Boggess}, A.,
  {Bowers}, C.~W., {Bruhweiler}, F.~C., {Clampin}, M., {Danks}, A.~C., {Green},
  R.~F., {Heap}, S.~R., {Hutchings}, J.~B., {Jenkins}, E.~B., {Joseph}, C.,
  {Kaiser}, M.~E., {Kimble}, R.~A., {Kraemer}, S., {Lindler}, D., {Linsky},
  J.~L., {Maran}, S.~P., {Moos}, H.~W., {Plait}, P., {Roesler}, F., {Timothy},
  J.~G., \& {Weistrop}, D. 2001, \aj, 122, 3396

\bibitem[{{Grady} {et~al.}(2005){Grady}, {Woodgate}, {Heap}, {Bowers}, {Nuth},
  {Herczeg}, \& {Hill}}]{grady2005}
{Grady}, C.~A., {Woodgate}, B., {Heap}, S.~R., {Bowers}, C., {Nuth}, III,
  J.~A., {Herczeg}, G.~J., \& {Hill}, H.~G.~M. 2005, \apj, 620, 470

\bibitem[{{Guimar{\~a}es} {et~al.}(2006){Guimar{\~a}es}, {Alencar}, {Corradi},
  \& {Vieira}}]{guimaraes2006}
{Guimar{\~a}es}, M.~M., {Alencar}, S.~H.~P., {Corradi}, W.~J.~B., \& {Vieira},
  S.~L.~A. 2006, \aap, 457, 581

\bibitem[{{Hashimoto} {et~al.}(2012){Hashimoto}, {Dong}, {Kudo}, {Honda},
  {McClure}, {Zhu}, {Muto}, {Wisniewski}, {Abe}, {Brandner}, {Brandt},
  {Carson}, {Egner}, {Feldt}, {Fukagawa}, {Goto}, {Grady}, {Guyon}, {Hayano},
  {Hayashi}, {Hayashi}, {Henning}, {Hodapp}, {Ishii}, {Iye}, {Janson},
  {Kandori}, {Knapp}, {Kusakabe}, {Kuzuhara}, {Kwon}, {Matsuo}, {Mayama},
  {McElwain}, {Miyama}, {Morino}, {Moro-Martin}, {Nishimura}, {Pyo}, {Serabyn},
  {Suenaga}, {Suto}, {Suzuki}, {Takahashi}, {Takami}, {Takato}, {Terada},
  {Thalmann}, {Tomono}, {Turner}, {Watanabe}, {Yamada}, {Takami}, {Usuda}, \&
  {Tamura}}]{hashimoto2012}
{Hashimoto}, J., {Dong}, R., {Kudo}, T., {Honda}, M., {McClure}, M.~K., {Zhu},
  Z., {Muto}, T., {Wisniewski}, J., {Abe}, L., {Brandner}, W., {Brandt}, T.,
  {Carson}, J., {Egner}, S., {Feldt}, M., {Fukagawa}, M., {Goto}, M., {Grady},
  C.~A., {Guyon}, O., {Hayano}, Y., {Hayashi}, M., {Hayashi}, S., {Henning},
  T., {Hodapp}, K., {Ishii}, M., {Iye}, M., {Janson}, M., {Kandori}, R.,
  {Knapp}, G., {Kusakabe}, N., {Kuzuhara}, M., {Kwon}, J., {Matsuo}, T.,
  {Mayama}, S., {McElwain}, M.~W., {Miyama}, S., {Morino}, J.-I.,
  {Moro-Martin}, A., {Nishimura}, T., {Pyo}, T.-S., {Serabyn}, G., {Suenaga},
  T., {Suto}, H., {Suzuki}, R., {Takahashi}, Y., {Takami}, M., {Takato}, N.,
  {Terada}, H., {Thalmann}, C., {Tomono}, D., {Turner}, E.~L., {Watanabe}, M.,
  {Yamada}, T., {Takami}, H., {Usuda}, T., \& {Tamura}, M. 2012, \apjl, 758,
  L19

\bibitem[{{Houk} \& {Cowley}(1975)}]{houk1975}
{Houk}, N. \& {Cowley}, A.~P. 1975, {University of Michigan Catalogue of
  two-dimensional spectral types for the HD stars. Volume I. Declinations -90\_
  to -53\_{f}0.}

\bibitem[{{Kenworthy} {et~al.}(2010){Kenworthy}, {Quanz}, {Meyer}, {Kasper},
  {Girard}, {Lenzen}, {Codona}, \& {Hinz}}]{kenworthy2010}
{Kenworthy}, M., {Quanz}, S., {Meyer}, M., {Kasper}, M., {Girard}, J.,
  {Lenzen}, R., {Codona}, J., \& {Hinz}, P. 2010, The Messenger, 141, 2

\bibitem[{{Liskowsky} {et~al.}(2012){Liskowsky}, {Brittain}, {Najita}, {Carr},
  {Doppmann}, \& {Troutman}}]{liskowsky2012}
{Liskowsky}, J.~P., {Brittain}, S.~D., {Najita}, J.~R., {Carr}, J.~S.,
  {Doppmann}, G.~W., \& {Troutman}, M.~R. 2012, \apj, 760, 153

\bibitem[{{Liu} {et~al.}(2003){Liu}, {Hinz}, {Meyer}, {Mamajek}, {Hoffmann}, \&
  {Hora}}]{liu2003}
{Liu}, W.~M., {Hinz}, P.~M., {Meyer}, M.~R., {Mamajek}, E.~E., {Hoffmann},
  W.~F., \& {Hora}, J.~L. 2003, \apjl, 598, L111

\bibitem[{{Mayama} {et~al.}(2012){Mayama}, {Hashimoto}, {Muto}, {Tsukagoshi},
  {Kusakabe}, {Kuzuhara}, {Takahashi}, {Kudo}, {Dong}, {Fukagawa}, {Takami},
  {Momose}, {Wisniewski}, {Follette}, {Abe}, {Akiyama}, {Brandner}, {Brandt},
  {Carson}, {Egner}, {Feldt}, {Goto}, {Grady}, {Guyon}, {Hayano}, {Hayashi},
  {Hayashi}, {Henning}, {Hodapp}, {Ishii}, {Iye}, {Janson}, {Kandori}, {Kwon},
  {Knapp}, {Matsuo}, {McElwain}, {Miyama}, {Morino}, {Moro-Martin},
  {Nishimura}, {Pyo}, {Serabyn}, {Suto}, {Suzuki}, {Takato}, {Terada},
  {Thalmann}, {Tomono}, {Turner}, {Watanabe}, {Yamada}, {Takami}, {Usuda}, \&
  {Tamura}}]{mayama2012}
{Mayama}, S., {Hashimoto}, J., {Muto}, T., {Tsukagoshi}, T., {Kusakabe}, N.,
  {Kuzuhara}, M., {Takahashi}, Y., {Kudo}, T., {Dong}, R., {Fukagawa}, M.,
  {Takami}, M., {Momose}, M., {Wisniewski}, J.~P., {Follette}, K., {Abe}, L.,
  {Akiyama}, E., {Brandner}, W., {Brandt}, T., {Carson}, J., {Egner}, S.,
  {Feldt}, M., {Goto}, M., {Grady}, C.~A., {Guyon}, O., {Hayano}, Y.,
  {Hayashi}, M., {Hayashi}, S., {Henning}, T., {Hodapp}, K.~W., {Ishii}, M.,
  {Iye}, M., {Janson}, M., {Kandori}, R., {Kwon}, J., {Knapp}, G.~R., {Matsuo},
  T., {McElwain}, M.~W., {Miyama}, S., {Morino}, J.-I., {Moro-Martin}, A.,
  {Nishimura}, T., {Pyo}, T.-S., {Serabyn}, E., {Suto}, H., {Suzuki}, R.,
  {Takato}, N., {Terada}, H., {Thalmann}, C., {Tomono}, D., {Turner}, E.~L.,
  {Watanabe}, M., {Yamada}, T., {Takami}, H., {Usuda}, T., \& {Tamura}, M.
  2012, \apjl, 760, L26

\bibitem[{{Mulders} {et~al.}(2013){Mulders}, {Paardekooper}, {Pani{\'c}},
  {Dominik}, {van Boekel}, \& {Ratzka}}]{mulders2013b}
{Mulders}, G.~D., {Paardekooper}, S.-J., {Pani{\'c}}, O., {Dominik}, C., {van
  Boekel}, R., \& {Ratzka}, T. 2013, \aap, 557, A68

\bibitem[{{Muto} {et~al.}(2012){Muto}, {Grady}, {Hashimoto}, {Fukagawa},
  {Hornbeck}, {Sitko}, {Russell}, {Werren}, {Cur{\'e}}, {Currie}, {Ohashi},
  {Okamoto}, {Momose}, {Honda}, {Inutsuka}, {Takeuchi}, {Dong}, {Abe},
  {Brandner}, {Brandt}, {Carson}, {Egner}, {Feldt}, {Fukue}, {Goto}, {Guyon},
  {Hayano}, {Hayashi}, {Hayashi}, {Henning}, {Hodapp}, {Ishii}, {Iye},
  {Janson}, {Kandori}, {Knapp}, {Kudo}, {Kusakabe}, {Kuzuhara}, {Matsuo},
  {Mayama}, {McElwain}, {Miyama}, {Morino}, {Moro-Martin}, {Nishimura}, {Pyo},
  {Serabyn}, {Suto}, {Suzuki}, {Takami}, {Takato}, {Terada}, {Thalmann},
  {Tomono}, {Turner}, {Watanabe}, {Wisniewski}, {Yamada}, {Takami}, {Usuda}, \&
  {Tamura}}]{muto2012}
{Muto}, T., {Grady}, C.~A., {Hashimoto}, J., {Fukagawa}, M., {Hornbeck}, J.~B.,
  {Sitko}, M., {Russell}, R., {Werren}, C., {Cur{\'e}}, M., {Currie}, T.,
  {Ohashi}, N., {Okamoto}, Y., {Momose}, M., {Honda}, M., {Inutsuka}, S.,
  {Takeuchi}, T., {Dong}, R., {Abe}, L., {Brandner}, W., {Brandt}, T.,
  {Carson}, J., {Egner}, S., {Feldt}, M., {Fukue}, T., {Goto}, M., {Guyon}, O.,
  {Hayano}, Y., {Hayashi}, M., {Hayashi}, S., {Henning}, T., {Hodapp}, K.~W.,
  {Ishii}, M., {Iye}, M., {Janson}, M., {Kandori}, R., {Knapp}, G.~R., {Kudo},
  T., {Kusakabe}, N., {Kuzuhara}, M., {Matsuo}, T., {Mayama}, S., {McElwain},
  M.~W., {Miyama}, S., {Morino}, J.-I., {Moro-Martin}, A., {Nishimura}, T.,
  {Pyo}, T.-S., {Serabyn}, E., {Suto}, H., {Suzuki}, R., {Takami}, M.,
  {Takato}, N., {Terada}, H., {Thalmann}, C., {Tomono}, D., {Turner}, E.~L.,
  {Watanabe}, M., {Wisniewski}, J.~P., {Yamada}, T., {Takami}, H., {Usuda}, T.,
  \& {Tamura}, M. 2012, \apjl, 748, L22

\bibitem[{{Panic} {et~al.}(2012){Panic}, {Ratzka}, {Mulders}, {Dominik}, {van
  Boekel}, {Henning}, {Jaffe}, \& {Min}}]{panic2012}
{Panic}, O., {Ratzka}, T., {Mulders}, G.~D., {Dominik}, C., {van Boekel}, R.,
  {Henning}, T., {Jaffe}, W., \& {Min}, M. 2012, ArXiv e-prints

\bibitem[{{Pantin} {et~al.}(2000){Pantin}, {Waelkens}, \&
  {Lagage}}]{pantin2000}
{Pantin}, E., {Waelkens}, C., \& {Lagage}, P.~O. 2000, \aap, 361, L9

\bibitem[{{Perrin} {et~al.}(2009){Perrin}, {Schneider}, {Duchene}, {Pinte},
  {Grady}, {Wisniewski}, \& {Hines}}]{perrin2009}
{Perrin}, M.~D., {Schneider}, G., {Duchene}, G., {Pinte}, C., {Grady}, C.~A.,
  {Wisniewski}, J.~P., \& {Hines}, D.~C. 2009, \apjl, 707, L132

\bibitem[{{Perryman} {et~al.}(1997){Perryman}, {Lindegren}, {Kovalevsky},
  {Hoeg}, {Bastian}, {Bernacca}, {Cr{\'e}z{\'e}}, {Donati}, {Grenon},
  {Grewing}, {van Leeuwen}, {van der Marel}, {Mignard}, {Murray}, {Le Poole},
  {Schrijver}, {Turon}, {Arenou}, {Froeschl{\'e}}, \&
  {Petersen}}]{perryman1997}
{Perryman}, M.~A.~C., {Lindegren}, L., {Kovalevsky}, J., {Hoeg}, E., {Bastian},
  U., {Bernacca}, P.~L., {Cr{\'e}z{\'e}}, M., {Donati}, F., {Grenon}, M.,
  {Grewing}, M., {van Leeuwen}, F., {van der Marel}, H., {Mignard}, F.,
  {Murray}, C.~A., {Le Poole}, R.~S., {Schrijver}, H., {Turon}, C., {Arenou},
  F., {Froeschl{\'e}}, M., \& {Petersen}, C.~S. 1997, \aap, 323, L49

\bibitem[{{Pineda} {et~al.}(2014){Pineda}, {Quanz}, {Meru}, {Mulders}, {Meyer},
  {Panic}, \& {Avenhaus}}]{pineda2014}
{Pineda}, J.~E., {Quanz}, S.~P., {Meru}, F., {Mulders}, G.~D., {Meyer}, M.~R.,
  {Panic}, O., \& {Avenhaus}, H. 2014, ArXiv e-prints

\bibitem[{{Pinte} {et~al.}(2008){Pinte}, {Padgett}, {M{\'e}nard},
  {Stapelfeldt}, {Schneider}, {Olofsson}, {Pani{\'c}}, {Augereau},
  {Duch{\^e}ne}, {Krist}, {Pontoppidan}, {Perrin}, {Grady}, {Kessler-Silacci},
  {van Dishoeck}, {Lommen}, {Silverstone}, {Hines}, {Wolf}, {Blake}, {Henning},
  \& {Stecklum}}]{pinte2008}
{Pinte}, C., {Padgett}, D.~L., {M{\'e}nard}, F., {Stapelfeldt}, K.~R.,
  {Schneider}, G., {Olofsson}, J., {Pani{\'c}}, O., {Augereau}, J.~C.,
  {Duch{\^e}ne}, G., {Krist}, J., {Pontoppidan}, K., {Perrin}, M.~D., {Grady},
  C.~A., {Kessler-Silacci}, J., {van Dishoeck}, E.~F., {Lommen}, D.,
  {Silverstone}, M., {Hines}, D.~C., {Wolf}, S., {Blake}, G.~A., {Henning}, T.,
  \& {Stecklum}, B. 2008, \aap, 489, 633

\bibitem[{{Quanz} {et~al.}(2013{\natexlab{a}}){Quanz}, {Amara}, {Meyer},
  {Kenworthy}, {Kasper}, \& {Girard}}]{quanz2013a}
{Quanz}, S.~P., {Amara}, A., {Meyer}, M.~R., {Kenworthy}, M.~A., {Kasper}, M.,
  \& {Girard}, J.~H. 2013{\natexlab{a}}, \apjl, 766, L1

\bibitem[{{Quanz} {et~al.}(2013{\natexlab{b}}){Quanz}, {Avenhaus}, {Buenzli},
  {Garufi}, {Schmid}, \& {Wolf}}]{quanz2013b}
{Quanz}, S.~P., {Avenhaus}, H., {Buenzli}, E., {Garufi}, A., {Schmid}, H.~M.,
  \& {Wolf}, S. 2013{\natexlab{b}}, \apjl, 766, L2

\bibitem[{{Quanz} {et~al.}(2011){Quanz}, {Schmid}, {Geissler}, {Meyer},
  {Henning}, {Brandner}, \& {Wolf}}]{quanz2011}
{Quanz}, S.~P., {Schmid}, H.~M., {Geissler}, K., {Meyer}, M.~R., {Henning}, T.,
  {Brandner}, W., \& {Wolf}, S. 2011, \apj, 738, 23

\bibitem[{{Schneider} {et~al.}(2003){Schneider}, {Wood}, {Silverstone},
  {Hines}, {Koerner}, {Whitney}, {Bjorkman}, \& {Lowrance}}]{schneider2003}
{Schneider}, G., {Wood}, K., {Silverstone}, M.~D., {Hines}, D.~C., {Koerner},
  D.~W., {Whitney}, B.~A., {Bjorkman}, J.~E., \& {Lowrance}, P.~J. 2003, \aj,
  125, 1467

\bibitem[{{Tatulli} {et~al.}(2011){Tatulli}, {Benisty}, {M{\'e}nard},
  {Varni{\`e}re}, {Martin-Za{\"i}di}, {Thi}, {Pinte}, {Massi}, {Weigelt},
  {Hofmann}, \& {Petrov}}]{tatulli2011}
{Tatulli}, E., {Benisty}, M., {M{\'e}nard}, F., {Varni{\`e}re}, P.,
  {Martin-Za{\"i}di}, C., {Thi}, W.-F., {Pinte}, C., {Massi}, F., {Weigelt},
  G., {Hofmann}, K.-H., \& {Petrov}, R.~G. 2011, \aap, 531, A1

\bibitem[{{van den Ancker} {et~al.}(1997){van den Ancker}, {The}, {Tjin A
  Djie}, {Catala}, {de Winter}, {Blondel}, \& {Waters}}]{vandenancker1997}
{van den Ancker}, M.~E., {The}, P.~S., {Tjin A Djie}, H.~R.~E., {Catala}, C.,
  {de Winter}, D., {Blondel}, P.~F.~C., \& {Waters}, L.~B.~F.~M. 1997, \aap,
  324, L33

\bibitem[{{van der Plas} {et~al.}(2009){van der Plas}, {van den Ancker},
  {Acke}, {Carmona}, {Dominik}, {Fedele}, \& {Waters}}]{vanderplas2009}
{van der Plas}, G., {van den Ancker}, M.~E., {Acke}, B., {Carmona}, A.,
  {Dominik}, C., {Fedele}, D., \& {Waters}, L.~B.~F.~M. 2009, \aap, 500, 1137

\bibitem[{{van Leeuwen}(2007)}]{vanleeuwen2007}
{van Leeuwen}, F. 2007, \aap, 474, 653

\end{thebibliography}

\end{document}